Determination of egg storage time at room temperature using a low-cost NIR spectrometer and machine learning techniques.

**Julian Coronel-Reyes[a], Ivan Ramirez-Morales*[ab], Enrique Fernandez-Blanco[b], Daniel Rivero[b], Alejandro Pazos[b]**

[a] **Universidad Técnica de Machala,** Faculty of Agricultural & Livestock Sciences. Address: 5.5 km Pan-American Av, Machala, El Oro, Ecuador.

[b] **Universidade A Coruña,** Department of Computer Science. Address: 15071 A Coruña (03082), A Coruña, España.

**Abstract**

Nowadays, consumers are more concerned about freshness and quality of food. Poultry egg storage time is a freshness and quality indicator in industrial and consumer applications, even though egg marking is not always required outside the European Union.

Other authors have already published works using expensive laboratory equipment in order to determine the storage time and freshness in eggs. Oppositely, this paper presents a novel method based on low-cost devices for rapid and non-destructive prediction of egg storage time at room temperature (23±1°C).

H&N brown flock with 49-week-old hens were used as source for the sampled eggs. Those samples were daily scanned with a low-cost smartphone-connected near infrared reflectance (NIR) spectrometer for a period of 22 days starting to run from the egg laid. The resulting dataset of 660 samples was randomly splitted according to a 10-fold cross validation in order to be used in a contrast and optimization process of two machine learning algorithms. During the optimization, several models were tested to develop a robust calibration model.

The best model use a Savitzky Golay preprocessing technique, and an Artificial Neural Network with ten neurons in one hidden layer. Regressing the storage time of the eggs, tests achieved a coefficient of determination (*R-squared*) of 0.8319±0.0377 and a root mean squared error (*RMSE*) of 1.97.

Although further work is needed, this technique has shown industrial potential and consumer utility to determine the egg's freshness by using a low-cost spectrometer connected to a smartphone.

# I. Introduction

Eggs are an affordable source of nutrients in the human diet. Its freshness and quality declines with the time, been influenced by the storage conditions. Degradation can get to the point of been unfitted for human consumption (Abdel-Nour, Ngadi, Prasher, & Karimi, 2011; Akter, Kasim, Omar, & Sazili, 2014; Akyurek & Okur, 2009; Mathew, Olufemi, & Foluke, 2016).

The variability in freshness might be perceived by consumers as lack of quality, for this reason it is important to study methods to monitor and preserve them better (Akter et al., 2014; Karoui et al., 2006).

Important and complex changes occur in egg during storage. Predicting these changes is critical in order to monitor egg freshness. These changes include thinning of albumin, weakening the vitelline membrane and increase the water content of the yolk. The properties of foaming and emulsifying the albumen and yolk, respectively, are affected by the protein concentration, pH, ionic strength (Karoui et al., 2006).

Storage time, temperature, humidity, air quality, and handling are external factors which causes the degradation of eggs. In particular, storage time is related with two major issues: the reduction of the nutritional value of eggs (Stadelman, Newkirk, & Newby, 1995) and the decreasing of freshness in a logarithmic relation (Silversides & Scott, 2001).

Akter et al (2014) demonstrated that egg weight, pH, oxidation and Haugh Units are also adversely affected with increasing storage time at room temperature. In the same work, the authors propose a maximum of 14 days for the time to be stored at room temperature.

The freshness can be assessed by physical, biochemical, microbial and sensory parameters. The Haugh Unit (HU) method is a widely used destructive method to measure egg quality (Haugh, 1937). However quality measurements based on HU are biased by the strain and age of the hen (Silversides & Scott, 2001). Lui et al (2007) demonstrated a high correlation between HU and storage time with a R-squared of 0.9868.

Sensor technologies are an attractive strategy for non-destructive determination of freshness of the egg, either at the production plant, or at food industries (Galiş, Dale, Boudry, & Thέwis, 2012; Karoui et al., 2006).

In the last years, non-destructive techniques for freshness and storage time at room temperature have emerged. This techniques include electronic nose (Yongwei, Wang, Zhou, & Lu, 2009), ultrasound (M. Aboonajmi, Setarehdan, Akram, Nishizu, & Kondo, 2014), ultraviolet-visible spectroscopy (Y. Liu et al., 2007) and near infrared spectroscopy (Abdel-Nour et al., 2011; Mohammad Aboonajmi, Saberi, Abbasian Najafabadi, & Kondo, 2015; Lin, Zhao, Sun, kun Bi, & Cai, 2015; Zhao et al., 2010).

The food industry has used NIR spectroscopy for long time (Stark, 1996) because it is an accurate, rapid, and non-destructive quality analysis technique (Kumaravelu & Gopal, 2015). Recent works have been published using NIR to predict storage time associated with its freshness, in atlantic salmon (Kimiya, Sivertsen, & Heia, 2013), large yellow croaker (Gangying et al., 2015), snow crab (Lorentzen, Rotabakk, Olsen, Skuland, & Siikavuopio, 2016/1), pork (Chen, Cai, Wan, & Zhao, 2011), apples (F. Liu & Tang, 2015), valerianella locusta (Giovenzana, Beghi, Buratti, Civelli, & Guidetti, 2014), and eggs (Abdel-Nour et al., 2011; Mohammad Aboonajmi et al., 2015; Lin et

al., 2015; Zhao et al., 2010).

In the past ten years, the evolution of small, hand-held instruments has seen considerable growth (Barton, 2016; Haughey, Galvin-King, Malechaux, & Elliott, 2014). Recently, some low-cost NIR devices have appeared in the market making NIRS applications affordable and therefore, accessible to a wider public (Haughey et al., 2014).

NIR spectra is the result of vibrational transitions associated with chemical bonds present in most organic compounds (dos Santos, Lopo, Páscoa, & Lopes, 2013; Kumaravelu & Gopal, 2015; Teye, Huang, & Afoakwa, 2013). The resulting spectrum is a consequence of the modifications made simultaneously in all the properties in the sample, making the calibration process more complicated (Florkowski, Prussia, Shewfelt, & Brueckner, 2009; Martens & Naes, 1992).

Chemometrics has become into an essential technique aimed to develop NIR calibration models. Using this techniques it is possible to process many numerous samples in a short time (Moros, Garrigues, & Guardia, 2010).

Multivariate analysis techniques are commonly used to process spectral data, techniques such as Principal Component Analysis (PCA), Partial Least Squares (PLS) have been widely used (Kumaravelu & Gopal, 2015). Recently, some machine learning techniques are popularizing as good alternatives to the classic techniques, since they are based on pattern recognition (Brereton, 2015).

The aim of this study was to assess the potential of a low-cost NIR spectrometer as a non-destructive and rapid technique for egg storage time assessment. More specific objective is to develop and evaluate a chemometric NIR calibration model based on machine learning techniques for the determination of egg storage time at room temperature.

## II. Materials and methods

Overall methodology, as can be viewed in Figure 1, consists of three moments: the acquisition of the data (Section 2.1), data partition using a cross-validation technique (Section 2.2) and the optimization of the chemometric model (Section 2.3). At each moment several steps were followed. In following subsections the methodology is described in detail.

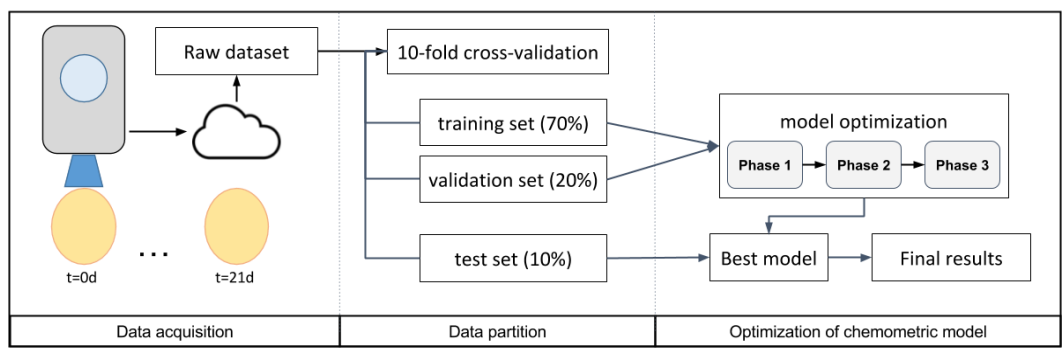

Figure 1. Diagram of moments and steps in experimental methodology

### 2.1 Data acquisition

Samples were collected using a smartphone-connected NIR spectrometer; each sample was uploaded to a cloud application using phone's internet to be stored in a dataset until the end of sample collection.

#### 2.1.1 Smartphone-connected NIR spectrometer

SCiO$^{TM}$ handheld NIR Spectrometer is a device with a built-in light source and a silicon sensor in a NIR short wavelength of 740nm to 1070nm (Goldring, Sharon, Brodetzki, & Ruf, 2016). The spectral data are transferred to a smartphone via Bluetooth and then to a cloud application (Goldring et al., 2016).

Previous works has already pointed to the potential of this low-cost device (Haughey et al., 2014; Schulte, Brink, Gruna, Herzog, & Gruger, 2015). This hardware allows to perform rapid tests which could be available on a smartphone (Cartwright, 2016; Das et al., 2015; Pügner, Knobbe, & Grüger, 2016).

It has also been reported the use of this device in research related to the detection of counterfeit medicines (Guillemain, Dégardin, & Roggo, 2017; Kaur, 2015; Wilson, Kaur, Allan, Lozama, & Bell, 2017) and for predicting the storage time and expiration packaged chicken fillets (Weesepoel & Ruth, 2016).

#### 2.1.2 Sample collection

A total of 660 spectral signals between 740nm to 1070nm were recorded with a spectral resolution of 1 nm. Spectral data were stored in a cloud based dataset with its corresponding reference values in the time storage.

Spectral curves correspond to 22 days of continuous monitoring, with 30 shell intact brown poultry eggs, with weights between 55g and 65 g. Eggs were collected from a flock of 20.000 hens of the strain H&N which were between 49-52 weeks old. Hens were housed in stacked cage system, and were fed with a standard ration without the use of laying egg promoters.

The spectral data used for experimentation was obtained by averaging two repeated measurements taken successively in the same place. Eggs were scanned in the poultry house immediately after been laid (day 0) and then transported to the laboratory in a thermally insulated container. Measures from day 1 to day 21 were obtained in laboratory conditions monitored hourly at 23±1°C and relative humidity 90±2%. The interval between each measurement was strictly 24 hours.

The procedure employed is simple, and the time required for each measurement is short. Non destructive technique was used in this experiment, since it was intended to understand how the spectrum is modified in each of the eggs over time.

Using a research license of SCiO Lab, egg spectral signals were downloaded and imported into Matlab (The MathWorks Inc., Natick, MA) in order to develop and optimize the chemometric models.

The dataset used for this experiments will be publicly available after manuscript is accepted. During the peer reviewing process, dataset is available for downloading in the following private link:
https://data.mendeley.com/datasets/6hn67h2trb/draft?a=76022a6d-e2f2-454a-b507-4c1b59d5d0c5

## 2.2 Data partition

Raw dataset was downloaded and then partitioned using a repeated 10-fold cross-validation technique in order to have training, validation and test subsets for the optimization of calibration model.

The model performance measures should be evaluated in a set of new data which are not been used for training the model. A good model should be able to make accurate estimations on this test data (Mucherino, Papajorgji, & Pardalos, 2009).

Cross-validation is one common technique applied in machine learning to maximize the use of available data. In this technique, dataset is randomly divided into multiple subsets for training and test the model. Cross-validation is used to avoid overfitting of the model. (Kuhn & Johnson, 2013; Refaeilzadeh, Tang, & Liu, 2009).

In this work, spectral data were divided into training (calibration), validation and test subsets using a variation known as repeated cross-validation (Garcia & Filzmoser, 2015; Kuhn & Johnson, 2013). A repeated 10-fold cross-validation technique was chosen. Therefore, data were splitted into 10 groups, which 9 are used as calibration and validation sets and the remaining one as a test set. This process was repeated 50 times.

The training and test set were changed until all folds have been tested. Data partition for each fold divided randomly the dataset, having 462 samples (70%) for training, 132 samples (20%) for validation and 66 samples (10%) for test.

## 2.3 Optimization of calibration model

The relationship between the response in the spectral region of the NIR spectra and the target is often a nonlinear type (Bertran et al., 1999). The origin of these nonlinearities is difficult to identify, for this reason, calibration is often performed using multivariate analysis (Martens & Naes, 1992). In order to develop a chemometric model, it is required the NIR spectra, the reference values for calibration and an algorithm to link them (Barton, 2016).

Parameters of the model were optimized in three consecutive phases. With the parameters achieved by the best model, there was performed an evaluation in unseen data (test set) in order to estimate future performance in new data.

This step was performed in three consecutive phases of optimization. In Phase 1, two modelling algorithms (Section 2.3.1 and 2.3.2) were tested simultaneously with seven preprocessing techniques (Section 2.3.3). In Phase 2, the parameters of selected model were tuned in order to optimize its performance. In Phase 3, the feature selection threshold (Section 2.3.4) was fine tuned.

### 2.3.1 Partial Least Squares (PLS)

PLS was introduced by the Swedish statistician Herman Wold (H. Wold, 1985). In chemometrics, PLS-regression is used as a basic method for relating two data matrices, by a linear multivariate model. However, this method goes beyond traditional regression since it has the ability to analyze data with incomplete, noisy and collinear variables (S. Wold, Sjöström, & Eriksson, 2001).

This method is widely applied on NIR spectroscopy where multiple input variables are required. The accuracy of the model in PLS-regression, improves when the number of relevant variables and the number of observations is increased (Hattori & Otsuka, 2017).

The PLS model is aimed to find in a multidimensional space, the direction in X, which explains the maximum variance direction in the Y. PLS regression is suited when the problem has more predictor variables than the number of observations, and when there could be multicollinearity among X values (Yu et al., 2017).

### 2.3.2 Artificial neural networks (ANN)

Artificial neural networks (ANN) are data-modeling tools aimed to analyze complex relationships between inputs and outputs. In recent years, ANN have become a subject of much relevance in the scientific and research field, they are inspired in the human central nervous system, which have lots of numerous cells that work quickly and help in decision making (Cascardi, Micelli, & Aiello, 2017).

The Multilayer Perceptron (MLP) is a type of layered neural network with connections between consecutive forwarding layers. Figure 2 shows the general scheme of a MLP, one input layer, one or more hidden layers and one output layer. The transfer function of neurons is commonly a sigmoid function, but other functions can also be used (Kruse et al., 2013; Ruck, Rogers, Kabrisky, Oxley, & Suter, 1990).

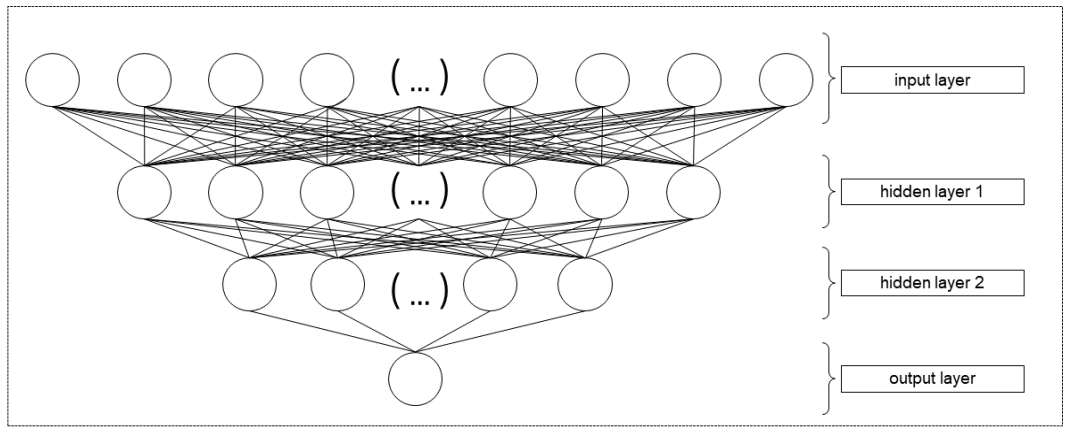

Figure 2. Multilayer perceptron representation.

Each neuron receives the output signals of the neurons in previous layer and provides an output for the next layer. The output layer receives as input the output of the last hidden layer and returns the output of the network. (Gardner & Dorling, 1998; Kruse et al., 2013).

The number of neurons, layers and their connections is commonly known as the architecture of the neural network, been one of the key parameters to be optimized. The architecture depends on the complexity of problem, and there is no general method for choosing the best one. Choosing a good architecture is an empirical process, where multiple architectures are tested in order to find one that offers satisfactory results (Herrera, Hervas, Otero, & Sánchez, 2004; Rivero, Fernandez-Blanco, Dorado, & Pazos, 2011).

The use of artificial neural networks has been successfully applied to NIR spectrometry in rapid quantification of wine compounds (Martelo-Vidal & Vázquez, 2015), evaluation of chemical components and properties of the jujube fruit (Guo, Ni, & Kokot, 2016), characterization of blends containing refined and extra virgin olive oils (Aroca-Santos, Cancilla, Pariente, & Torrecilla, 2016).

### 2.3.3 Preprocessing techniques

NIR spectra can be preprocessed to eliminate physical phenomena which alter raw spectra. This practice is an integral part of the chemometric modeling (Blanco & Villarroya, 2002) and it is considered one of the most important steps. The data contained on NIR spectra is the composition of several signals with overlapping information. (Blanco & Villarroya, 2002; dos Santos et al., 2013; Rinnan, Berg, & Engelsen, 2009).

Proper choice of technique preprocessing is difficult to assess before the validation of the model, therefore NIR spectra preprocessing is performed by trial and error (Rinnan et al., 2009; Xu et al., 2008).

In this work, raw spectral signal and other six common preprocessing techniques are analyzed:

**1) Raw spectra:** NIR-reflectance measurement of a sample, includes both the diffusively reflected and specular reflected radiation. The latter as does not contain relevant information and it is commonly minimized by instrument design and sampling geometry. The diffusively reflected light, is the primary source of information in the NIR spectra. (Rinnan et al., 2009).

**2) Savitzky Golay:** This pre-processing technique was proposed in 1964 by the authors who gave name to the technical (Savitzky & Golay, 1964). The authors popularized a method which includes a smoothing step for numerical derivation of a vector. In this method a p-order polynomial is fitted in a symmetric window of w-width on the raw data, then the d-order derivative is calculated at centre point i.

This operation is applied sequentially to all spectral points. The width of the window size, the degree of the fitted polynomial, are decisions that need to be made. The highest derivative that can be determined is the degree of the polynomial (Rinnan et al., 2009).

For initial experiments the width was se to five, and both orders of polynomial and derivative were set to two. In subsequent experiments, the width was modified from 3 to 101 (odd numbers only), the polynomial order from one to five, and the derivative order from one to five.

**3) Beer-Lambert law:** suggests a linear relationship between the reflectance of the spectra and concentration of components, the law can be expressed as shown in Equation 1 (Rinnan et al., 2009):

$$A\lambda = -\log_{10}(R) \cong \epsilon\lambda \times l \times c \quad (1)$$

Where $A\lambda$ is the wavelength-dependent absorbance, R is the reflectance, $\epsilon\lambda$ is the wavelength-dependent molar absorptivity, l is the length of the light through the sample matrix and c is the concentration of the components of interest.

**4) Standard Normal Variate (SNV):** is a method for scattering correction of NIR data (Barnes, Dhanoa, & Lister, 1989). The formula is expressed in Equation 2.

$$X_{i,snv} = \frac{X_i - \bar{x}}{S} \quad (2)$$

Where $X_{i,snv}$ is the SNV at a wavelength *i*, $\bar{x}$ is the spectrum average of the sample to be corrected, and *S* is the standard deviation of the spectrum sample.

**5) Multiplicative Scatter Correction:** is a widely used pre-processing technique for NIR, it was first introduced by Martens et al. (1983). In this technique, undesirable scatter effects of spectra are removed from the data matrix prior to data modeling.

MSC comprises in a first step the estimation of the correction coefficients, and a second step which consists in the correction of the recorded spectrum. Equation 3 and 4 show both steps respectively.

$$X_o = b_o + b_{r,1} \times X_r + e \quad (3)$$

$$X_c = \frac{X_o - b_o}{b_{r,1}} = X_r + \frac{e}{b_{r,1}} \quad (4)$$

Where $X_c$ is the corrected spectra, $X_o$ is one original sample spectra measured by the NIR instrument, $X_r$ is a reference spectrum used for preprocessing of the entire dataset. In most applications, the average spectrum of the calibration set is used as the reference spectrum. $e$ is the unmodeled part of $X_o$. $b_o$ and $b_{r,1}$ are the scalar parameters, which differ for each sample. (Rinnan et al., 2009).

**6) First Spectral Derivative (FSD):** these technique have been used in analytical spectroscopy for decades due to its ability to remove additive and multiplicative effects of the spectra (dos Santos et al., 2013). Finite difference is the basic method for spectral derivation (Rinnan et al., 2009); thus, the first derivative is calculated as the difference between two subsequent spectral points, as can be analyzed in Equation 5.

$$X_{i,fsd} = X_i - X_{i-1} \quad (5)$$

Where $X_{i,fsd}$ denotes the first derivative in the wavelength i, and $X_{i,ssd}$ represents the second derivative in the wavelength i. The first derivative removes the baseline of spectra.

**7) Second Spectral Derivative (SSD):** This technique is also based on finite difference for spectral derivation (Rinnan et al., 2009); thus, the second derivative is calculated as the difference between two subsequent points from processed FSD signal. The formula is shown in Equation 6 :

$$X_{i,ssd} = X_{i,fsd} - X_{i,fsd-1} \quad (6)$$

Where $X_{i,ssd}$ represents the second derivative in the wavelength i. The second derivative besides removing the baseline, remove the linear trend of spectra.

According to Rinnan et al., (2009) this technique should be avoided in practice, since it is not feasible for most real measurements due to noise inflation.

**2.3.4 Feature selection threshold**

Commonly NIR spectra contains a large amount of information along the wavelength range. For this reason, it is important to perform a technique aimed to reduce this amount of data (Blanco & Villarroya, 2002). This techniques have now become a necessity and a requirement in chemometrics (Guyon, Gunn, Nikravesh, & Zadeh, 2008; Saeys, Inza, & Larrañaga, 2007).

In machine learning techniques, the proper selection of features to get of a small subset with lower sensitivity to non-linearities is usually effective to improve performance of the models (Leardi, Boggia, & Terrile, 1992; Saeys et al., 2007).

Selection of relevant features related to the compound of interest and avoiding interference of others should be a target in terms of building a robust predictive model. In this study, a feature selection method was applied to select the informative wavelengths.

Feature Selection does not alter the original representation of the variables. This techniques simply choose a subset of the best wavelengths for the model (Saeys et al., 2007). A threshold is normally used together with a filter technique either univariate or multivariate model to evaluate which are the best wavelengths (Szymańska et al., 2015).

The present work used a multivariate filter described by Hall (1999) called Correlation-based Feature Selection (CFS). This filter is a simple algorithm that ranks feature importance according to its correlation function with the predicted variable. By using this method, the relevant informative wavelengths will be selected and therefore, it is expected to obtain an improvement of the model's performance.

**2.3.5 Model evaluation**

All models were evaluated using performance measures as the coefficient of determination (R-squared) in validation set from a cross-validation (dos Santos et al., 2013; Viscarra Rossel, 2008).

Mean and standard of R-squared obtained of 500 models (50 repetitions by 10-fold cross-validation) were calculated for each model configuration in order to decide the best parameters of the model.

Final results are presented in test set from a cross-validation. R-squared, mean absolute error (MAE) and root mean square error (RMSE) were obtained according to equations 7-9, respectively. MAE and RMSE, represent the difference between predictive values and the actual values (Armstrong & Collopy, 1992; Hyndman & Koehler, 2006).

$$R-squared = \frac{\sum_{i=1}^{i=n}(y_i - \hat{y}_i)^2}{\Box} \quad (7)$$

$$MAE = \frac{\sum_{i=1}^{i=n} |y_i - \hat{y}_i|}{\Box} \quad (8)$$

$$RMSE = \sqrt{\Box} \quad (9)$$

Where $y_i$ is the real value of the $i$-th observation, $\hat{y}_i$ is the predicted value of the $i$-th observation, $\bar{y}$ is the average of real values and $n$ is the number of observations.

**III. Results and Discussion**

This study presents a method to estimate the egg's storage time at room temperature using a smartphone connected NIR spectrometer and machine learning techniques. Data acquisition and data partition were broadly covered in section 2.1 and 2.2. Therefore in this section results obtained in the three phases of the calibration model optimization are presented.

**3.1 Phase 1: Selection of modelling algorithm and pre-processing technique**

This phase was aimed to simultaneously choose the modelling algorithm and the pre-processing techniques. A grid search technique (Koch et al., 2012; Ma, Zhang, & Wang, 2015), was used to evaluate models.

PLS and ANN algorithms were trained using the same partition schema to ensure that both models receive exactly the same input data, making the results comparable. Due to the influence of the feature selection threshold, the preprocessing techniques were evaluated at all thresholds using a correlation based feature selection (Hall, 1999).

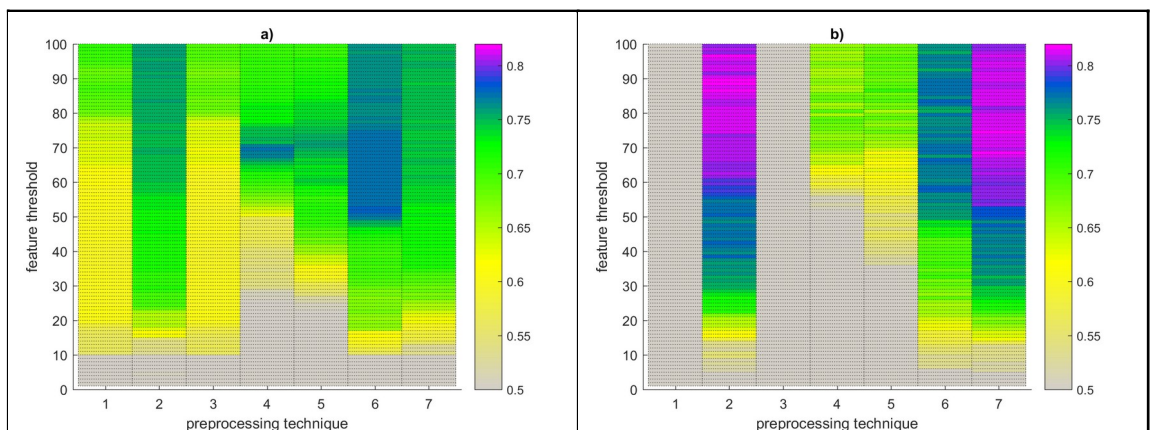

Figure 3: R-squared in CV validation set obtained with the seven preprocessing techniques at different values of feature threshold. a) Using PLS b) using ANN.

The results of the mean R-squared of 50 repeated 10-fold cross-validation achieved using PLS with 10 latent variables, can be seen in Figure 3a. Best results are obtained with pre-processing technique 4 (Standard Normal Variate) at 70% feature selection threshold and pre-processing technique 6 (First Spectral Derivative) at 50% feature selection threshold. In both cases, the results are below 0.8.

In Figure 3b, can be observed the results of the mean R-squared of 50 repeated 10-fold cross-validation achieved using ANN with 10 neurons in one hidden layer. Best results are obtained with pre-processing technique 2 (Savitzky Golay) at 90% feature selection threshold, and pre-processing technique 7 (Second Spectral Derivative) at 60% feature selection threshold. In both cases, the R-squared results are above 0.8. ANN models outperform to the PLS models. Therefore, ANN will be used to be optimized.

Despite the fact Savitzky Golay and Second Spectral Derivative have similar results, the latter one should be avoided in practice, because it is not feasible in practice due to the added noise (Rinnan et al., 2009). Additionally, Savitzky Golay technique, has tuning parameters which can be useful to optimize the model (Savitzky & Golay, 1964).

Therefore, once the preprocessing was chosen, the optimization was executed again to optimize the feature selection threshold again. Main reason is Savitzky-Golay's tuning parameters can transform the input pattern, and thus, a reevaluation of the selected wavelengths is necessary. Therefore, this parameter is set to 100% for now.

Savitzky Golay widths, odd numbers between 3 and 101 were tested for different configurations of Polynomial order and Derivative order (between first and fifth). Table 1 show the results of the mean, standard deviation, min and max, and p-value of a Tukey Honest Significant Difference (Tukey, 1949) test of R-squared of 50 repeated 10-fold cross-validation achieved at the best width of each configuration.

There is an intrinsic redundancy in the hierarchy of Savitzky Golay derivation. For each Polynomial order, the subsequent derivative order, gave the same estimate of the coefficients. In example, for the first-degree polynomial, a first derivative and second derivative will give the same answer. The results of redundant configurations are not been presented.

Table 1. Results of SavGol parameters at best width, using odd numbers from 3 al 101

| Polynomial degree | Derivative order | width | mean ± std | min | max | Tukey HSD |
|---|---|---|---|---|---|---|
| First | First | 7 | 0,7644 ± 0,069 | 0,360 | 0,861 | c |
| Second | First | 7 | 0,7654 ± 0,067 | 0,445 | 0,894 | c |
| Second | Second | 29 | 0,8204 ± 0,047 | 0,610 | 0,909 | ab |
| Third | First | 19 | 0,7678 ± 0,065 | 0,472 | 0,880 | c |
| Third | Second | 21 | 0,8214 ± 0,045 | 0,675 | 0,910 | ab |
| Third | Third | 53 | 0,8249 ± 0,044 | 0,690 | 0,926 | ab |
| Fourth | First | 13 | 0,7679 ± 0,073 | 0,267 | 0,899 | c |
| Fourth | Second | 41 | 0,8222 ± 0,045 | 0,561 | 0,926 | ab |
| Fourth | Third | 51 | 0,8248 ± 0,043 | 0,618 | 0,914 | ab |
| Fourth | Fourth | 61 | 0,8199 ± 0,042 | 0,651 | 0,909 | ab |
| Fifth | First | 5 | 0,7684 ± 0,07 | 0,123 | 0,889 | c |
| Fifth | Second | 39 | 0,8252 ± 0,045 | 0,624 | 0,931 | a |
| Fifth | Third | 67 | 0,8276 ± 0,041 | 0,590 | 0,910 | a |
| Fifth | Fourth | 61 | 0,8189 ± 0,042 | 0,660 | 0,907 | ab |
| Fifth | Fifth | 67 | 0,8107 ± 0,045 | 0,607 | 0,913 | b |

*Rows with different letters differ significantly according to Tukey HSD for a value of p <0.01.*

According to the Tukey HSD test, best configuration of Savitzky Golay can be a fifth polynomial degree and a second or third derivative order. The latter was chosen since it has a greater mean and smaller standard deviation in test data.

Therefore the optimized Savitzky Golay technique for this experiment was of 67 smoothing points (width), a fifth polynomial degree and a third order derivative.

There were evident differences in the spectra of the eggs as a function of storage time. Figure 4a shows the result of applying Savitzky Golay preprocessing technique. As the eggs were stored for longer, the obtained spectra takes a different characteristic values. This differences can also be seen in Figure 4b, which was made with an average of the signals corresponding to the eggs stored at 0, 7, 14 and 21 days.

Visually exploring the dataset it is found some patterns indicating that spectra of eggs changes as the storage time is increased. Notice in Figure 4b at wavelengths of 860-940 nm, that fresh eggs with zero days of storage have a Savitzky Golay reflectance near to zero, while stored eggs have oscillations between 1 and -1 in the same wavelength. After this pattern fresh eggs have a peak at 933nm, while stored eggs show a similar peak earlier at 913nm.

At 983nm and 999nm, the difference between eggs stored during 7 days and those stored more time is evident. In a visual exploration of data it is difficult to find differences between eggs stored for more than 14 days. According to Akter et al (2014) eggs must be stored maximum of 14 days at room temperature. The small differences between spectra of egg stored for more than 14 days may be a indicative of a deterioration in freshness of eggs.

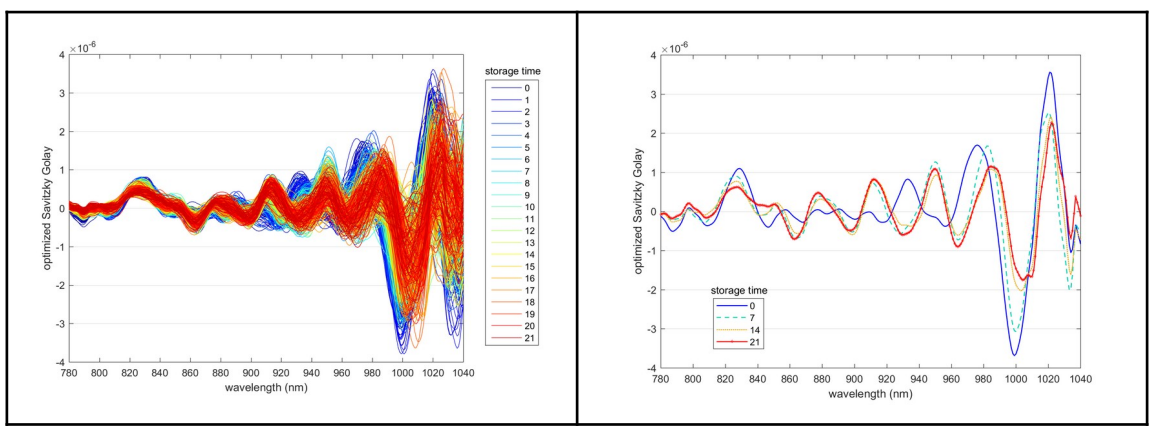

Figure 4. Spectral signal processed with optimized Savitzky Golay at different storage times. 4a) all spectral signals. 4b) mean spectra at 0,7,14,and 21 days of storage.

### 3.2 Phase 2: Optimization of selected algorithm parameters

During phase 1, the model using an ANN algorithm obtained better results. For this reason, ANN algorithm was selected. In this phase, it was decided to optimize the architecture of the neural network. Architectures of one and two layer were evaluated using a grid search (Koch et al., 2012; Ma et al., 2015). Values ranging from 0 to 200 neurons in each layer with intervals of 10 were evaluated.

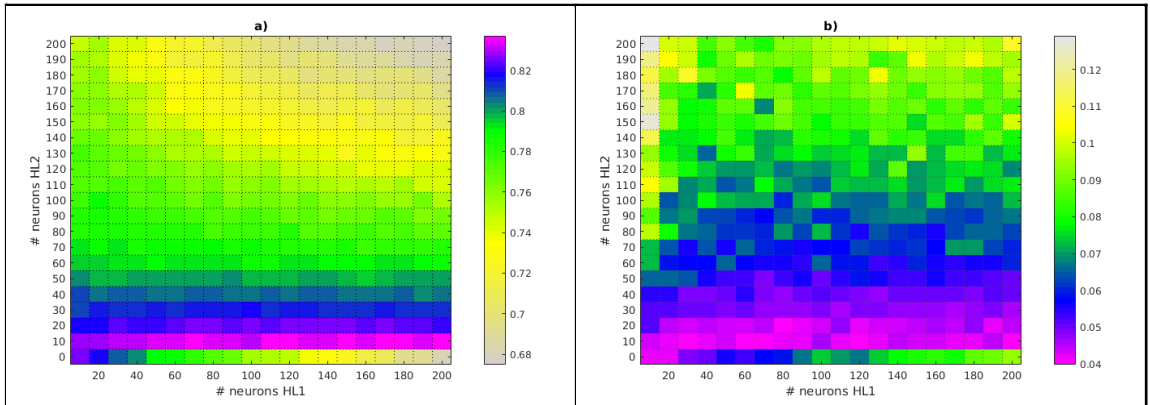

Figure 6. R-squared in 50 repeated 10-fold CV validation set. a) mean b) standard deviation

Figure 6b shows the standard deviation of 50 repeated 10 fold cross-validated models at each architecture, this measure indicates the stability and the repeatability of the proposed model. Best results are achieved in architectures of one hidden layer with 10 or 20 neurons, and architectures of two hidden layers in which the second layer has 10 or 20 neurons.

Although the best results are achieved with an architecture of two hidden layers with 180 and 10 neurons respectively. It was observed that using an architecture of 10 neurons in one hidden layer the model obtains statistically equal results. According to Abu-Mostafa (1989) in this work is has been selected the least complex with the same results, this is an architecture of 10 neurons in one hidden layer.

### 3.3 Phase 3: Fine tuning of feature selection threshold

The following experiment aims to find the best threshold value for feature (75, 150) selection. Figure 5 shows the influence of feature selection threshold in R-squared results.

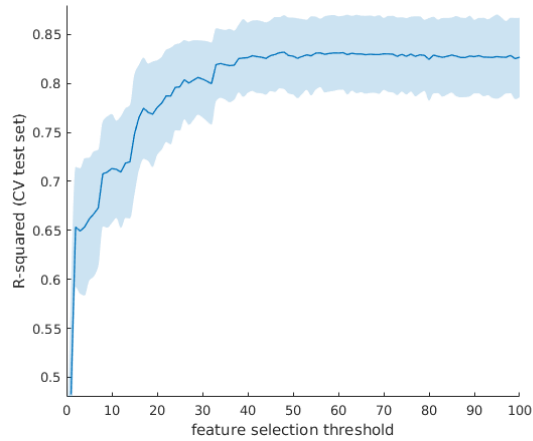

Figure 5. Results of R-squared at different values of feature threshold

Results above 0.8 are obtained with values of feature selection threshold from 30 onwards. Figure 5 evidences the importance of the appropriate selection of features as inputs of the model. A feature selection threshold of 48 is selected since R-squared obtained in test data is equal to 0.8319 ± 0.0377.

Although, evaluation of storage time of poultry eggs involves a complex process, this study shows that egg quality can be predicted using a ANN model with relevant features as input patterns. Our results confirm the stated by Guyon et al. (2008), that appropriate selection of relevant wavelengths is very important and must be used as a good strategy to avoid the inclusion of uninformative and redundant wavelengths in the predictive model.

**3.4 Evaluation of the model in unseen data (test set).**

Once optimized all parameters, final configuration of the model is a Savitzky-Golay pre-processing technique, with a fifth polynomial degree, a third derivative order and a width of 67. The ANN model has 10 neurons in one hidden layer and the feature selection threshold is of 48.

The calibration model using the previous mentioned parameters was tested in unseen data to evaluate its performance and generalization ability.

Figure 6a shows a regression plot between the values of egg storage time predicted by the model, with their corresponding actual valors. It can be seen that most predictions present an absolute error of about 2 days, some predictions show a error of about 4 days, and one value has the worst error of 6 days. Figure 6b shows the histogram of absolute errors.

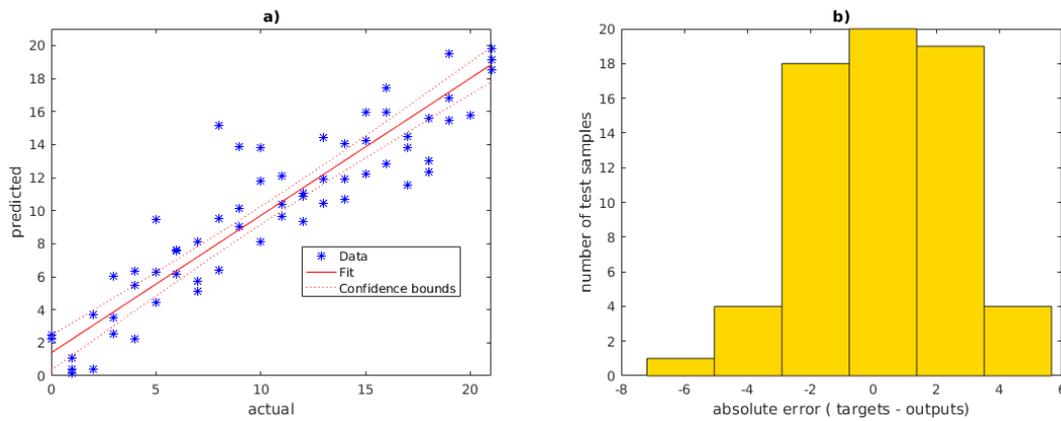

Figure 6. Model performance in test set data. a) Regression fit plot actual vs predicted. b) Absolute error histogram.

In order to evaluate the performance of the model in unseen data, R-squared and RMSE was calculated for the model and can be seen in Table 2. Proposed model has a high significance in all subsets.

Table 2. Performance metrics in training, validation y test set

|  | training | validation | test |
|---|---|---|---|
| **n samples** | 462 | 132 | 66 |
| **R-squared** | 0.865 | 0.862 | 0.873 |
| **Mean Absolute Error** | 1.834 | 1.998 | 1.810 |
| **Root Mean Squared Error** | 2.15 | 2.19 | 1.97 |
| **F-statistic vs. constant model** | 3.09e+03 | 673 | 449 |
| **p-value** | <0.01 | <0.01 | <0.01 |

The differences between results in test and training set performance could indicate the degree of overfitting (Stockwell & Peterson, 2002). However, Table 2 shows that these parameters are similar between training set, validation set and test set. Therefore we can affirm that the models are well fitted within the context of the calibration set.

Although the R-squared as a performance metric of the model is below 0.9, the RMSE is about two days, which gives safety margins to the user.

Some authors have evaluated predictive models for freshness assessment. Lin et. al. (2011) obtained a R-squared of 0.879 and a RMSE 2.443 using NIR spectroscopy for Haugh unit. Sun et. al. (2015) used artificial vision and dynamic weighing, to obtain a R-squared of 0.8653 and a RMSE of 3.745. Abdel-Nour et. al. (2011) achieved a R-squared of 0.89 and a RMSE of 1.65 in validation dataset for the prediction of storage days, using a lab grade VIS/NIR spectroradiometer.

Our results are similar to the mentioned above and this clearly indicate that there a is potential for portable NIR instruments as an alternative to desktop lab instrumentation in agree with Haughey (2014). Moreover, due to the recent advances in sensors technology such as Changhong (Rateni, Dario, & Cavallo, 2017) which have a embedded NIR sensor. It is possible to create applications using the proposed model. This approach could enable the consumers to predict the storage time in poultry eggs.

**IV. Conclusions and future work**

Our findings show the potential of smartphone-connected devices based on short wave NIR as an effective method for the evaluation of storage time in poultry eggs. Spectral analysis eggs is a rapid and nondestructive method for egg storage time determination.

Reflectance spectral data of the egg contains relevant information about its storage time, a parameter of freshness. There were noticeable differences in the processed spectra values of the eggs as a function of storage time.

Suitable predictive models were built with PLS and ANN regression techniques, however the latter performed better achieving a R-squared of 0.873 and a RMSE of 1.97 in test set data, suggesting that the spectra obtained with the smartphone connected NIR spectrometer can be used as a nondestructive method for the assessment of egg storage time, a parameter of quality and freshness.

The use of a smartphone connected NIR spectrometer is recommended for egg storage time assessment. However, further work is needed to assess the long-term reliability of the system. A combination of this method with destructive techniques is recommended.

Future work focus on the exhaustive experimentation at room and refrigeration temperatures using this low-cost spectrometer on eggs from hens of diverse ages and strains. The use of hyperspectral imaging for storage time prediction is also a potential technique to be studied.

**Acknowledgements**
The authors wish to thank Agrolomas CL for providing samples directly from their poultry houses. An special acknowledgement to CEDIA National Research and Education Network. Iván Ramírez-Morales and Enrique Fernández-Blanco would also like to thank the support provided by the NVIDIA Research Grants Program. This work is part of DINTA-UTMACH and RNASA-UDC research groups.